\RequirePackage{fix-cm}
\documentclass{svjour3}                     % onecolumn (standard format)
\smartqed  % flush right qed marks, e.g. at end of proof
\usepackage{graphicx}
\begin{document}

\title{Spectral study of GX 339-4 with TCAF using Swift/XRT and NuSTAR Observation}
%\subtitle{Spectral study of GX 339-4 with TCAF}

\titlerunning{Spectral study of GX 339-4 with TCAF}        % if too long for running head

\author{Santanu Mondal \and Sandip K. Chakrabarti         \and
        Dipak Debnath %etc.
}


\institute{S. Mondal \at
              Instituto de F\'isica y Astronom\'ia, Facultad de Ciencias, Universidad de Valpara\'iso, Gran Bretana N 1111, Playa Ancha, Valparaíso, Chile\\
              Indian Centre For Space Physics, 43 Chalantika, Garia Station Road, Kolkata 700084, India\\
%              Fax: +123-45-678910\\
              \email{santanu.mondal@uv.cl}           %  \\
           \and
           S. K. Chakrabarti \at
              S.N. Bose National Center for Basic Sciences, JD-Block, Salt Lake, Kolkata,700098, India
              Indian Centre For Space Physics, 43 Chalantika, Garia Station Road, Kolkata 700084, India
	   \and 
	   D. Debnath \at
	  Indian Centre For Space Physics, 43 Chalantika, Garia Station Road, Kolkata 700084, India
}

\date{Received: date / Accepted: date}

\maketitle

\begin{abstract}
We fit spectra of galactic transient source GX~339-4 during its 2013 outburst using
Two Component Advective Flow (TCAF) solution. For the first time, we are fitting combined NuSTAR 
and Swift observation with TCAF. We use TCAF to fit 0.8-9.0~keV Swift and 4-79 keV NuSTAR 
spectra along with the LAOR model. To fit the 
data we use disk accretion rate, halo accretion rate, size of the Compton cloud and the density jump 
of advective flows at this cloud boundary as model parameters. From TCAF fitted flow parameters, and 
energy spectral index we conclude that the source was in the hard state throughout this 
particular outburst. The present analysis also gives some idea about the broadening of 
Fe $K_{\alpha}$ with the accretion rate. Since TCAF does not include Fe line yet, we make use of 
the `LAOR model' as a phenomenological model and find an estimate of the Kerr parameter 
to be $\sim 0.99$ for this candidate.
\end{abstract}

\section{Introduction}
\label{intro}
Transient black hole binaries are ideal objects to study accretion physics. 
They spend most of the times in the quiescence state and occasionally produce outbursts. 
Their spectra have mainly two components: a multi-colored black body and a power-law component. 
The hot Compton cloud \cite{st80,hm93,zdziarski03} %(Sunyaev \& Titarchuk 1980; Haardt \& Maraschi 1993; Zdziarski et al. 2003) 
close to the black hole reprocesses soft photons from the disk and produces a power-law tail. 
Two component advective flow (TCAF) is a self-consistent 
solution of a viscous transonic flow in presence of radiative transfer. 
It was shown that \cite{c90} flows above a critical viscosity parameter $\alpha = \alpha_{crit}$ will 
become a Keplerian disk \cite{ss73}, while those below will remain sub-Keplerian. 
The sub-Keperian halo creates a high density centrifugal barrier
or even passes through a shock close to the black hole. This hot, puffed up region acts as the Compton cloud. 
Two accretion rates control spectral states of the black hole: 
if the accretion rate of the disk is large enough to cool down 
the CENtrifugal pressure supported BOundary Layer (CENBOL) which is the post-shock region, 
one obtains a soft state. The opposite is true for the hard state.
The spectral properties of TCAF is studied in \cite{ct95}
and \cite{c97}. In certain situations this CENtrifugal pressure dominated BOundary Layer or CENBOL 
exhibits radial and vertical oscillations and thus the Comptonized photon number also oscillates
and QPOs are produced. This region is also responsible to produce outflows. A recent review on TCAF
could be viewed in \cite{c015}.  

A large number of works attempted to explain evolution of spectral states \cite{vdk04,rm06,motta09}
%(van der Klis 2004; Remillard \& McClintock 2006; Motta et al. 2009) 
during outbursts of transient BHCs having temporal and spectral states 
with phenomenological models, such as disk black-body (DBB), power-law (PL), etc. 
However, using TCAF, state transitions and the origin of quasi-periodic oscillations (QPOs) have been 
successfully modelled by our group \cite{d13,nandi12}. %(Debnath et al. 2013; Nandi et al. 2012). 
In TCAF, oscillations of the post-shock region is either due to the fact that the resonance condition 
\cite{msc96,c15} %(Molteni et al. 1996; Chakrabarti et al. 2015) 
is satisfied, or the Rankine-Hugoniot condition is not satisfied \cite{rcm97}. %(Ryu et al. 1997). 
For transient sources, QPOs evolved monotonically \cite{c08,d10,d13}
%(Chakrabarti et al. 2008; Debnath et al. 2010, 2013)
due to the presence of Compton cooling \cite{mc13} %(Mondal \& Chakrabarti, 2013) 
and variation of viscosity \cite{mcd15}. %(Mondal et al. 2015). 
It rises during the rising phase and decreases in the declining
phase. After the inclusion of TCAF (\cite{dcm14,dd15a,dd15b,mdc14,molla16,jana16,chat16}) 
%(Debnath et al. 2014, 2015a, 2015b; Mondal et al. 2014; Molla et al. 2016; 
%Jana et al. 2016; Chatterjee et al. 2016) 
as a local additive Table model in HEASARC's spectral analysis software
package XSPEC, physical understanding behind the variation of spectral states in an outburst became easier. 
TCAF model {\it fits} is found to be an effective tool to predict dominating (type `C') QPOs 
directly from the spectral fitted shock parameters. Mass measurement (\cite{molla16,jana16}) %(Molla et al. 2016; Jana et al. 2016) 
becomes possible from the consideration of constant normalization factor during the
fitting of observed data using TCAF. These encouraging aspects motivated us to fit the
data with TCAF, even when they may also be fitted with other phenomenological models (e.g., DBB plus PL).  

GX 339-4 was first observed in 1973 \cite{markert73} %(Markert et al. 1973) 
by $1-60$ keV MIT X-ray detector onboard OSO-7 satellite. 
This stellar-mass black-hole binary has a mass function of $M_{bh}~sin(i)$ = $5.8\pm0.5~M_\odot$ and 
low-mass companion of mass $m = 0.52~M_\odot$ \cite{hynes03}. We adopt this value throughout the paper. 
This binary system is located at a distance of $d~\geq~6$~kpc \cite{hynes03} %(Hynes et al. 2003) 
with R.A.=$17^h02^m49^s.56$ and Dec.=$-48^\circ46'59''.88$. 
As the GX~339-4 is a non-eclipsing binary, the inclination angle should be less 
than $60^o$ \cite{cowley02}. Estimation by \cite{zdziarski04} %Zdziarski et al. (2004) 
gives a lower limit $45^o$ from the secondary mass function. 
Joint modeling of Suzaku and XMM-Newton observation was 
used by \cite{miller08} to explain relativistic broadening of Fe~$K_\alpha$ line 
and obtain the spin parameter to be $a = 0.93\pm0.01$. Using RXTE and XMM-Newton data,
\cite{reis08} observed a distorted Fe~$K_\alpha$ line above the continuum. 
Recently, \cite{plant14} calculated the value of spin parameter $a = 0.9$ with 
$33^o$ inclination angle with reflection and disk continuum modeling. Thus the exact values of 
the disk inclination angle and spin parameter are still doubtful. This required observations through more sensitive 
and high resolution instruments.

There are extensive discussions on temporal, spectral and multi-wavelength properties of the source during 
its different outbursts \cite{nandi12,dincer12} using RXTE/PCA observation. 
Some temporal and spectral properties of this candidate during its 2013 outburst have been discussed
by \cite{bachetti15} and \cite{furst15} using NuSTAR and Swift \cite{gehrel04} 
observations with truncated disk model. The authors reported the presence of an Iron line.
There are many recent analysis on this object in the literature regarding spectral and timing properties
\cite{bz16}. %(Basak \& Zdziarski, 2016 and references therein).
In this {\it paper}, for the first time, TCAF model fitting procedure was applied for the combined
NuSTAR \cite{harrison13} and Swift observations. Since TCAF represents solutions of governing equations
which include radiation mechanism, we expect that fits would yield actual physical parameters such as the 
accretion rates of two components and the size and other properties of the Compton cloud on the 
days of the observation. Specifically we want to check whether TCAF also requires an iron line and if so,
whether it would be broad double peaked iron line as observed in Suzaku and XMM-Newton observations 
\cite{miller06,brenneman06,tomsick08}. %(Miller et al. 2006; Brenneman \& Reynolds 2006; Tomsick et al. 2008).
Earlier, \cite{abell79} described very accurately the splitting of red/blue shifted lines 
from their jet kinematic model (so-called ballistic motion) in SS433. 
Line profile for Schwarzschild candidates was studied by \cite{fabian89}. %Fabian et al. (1989). 

The {\it paper} is organized in the following way: in the next Section, we briefly discuss observation
and data analysis procedures using HEASoft and NuSTARDAS software. 
In \S 3, we present fitted results obtained from TCAF solution of Swift and NuSTAR data. We also discuss the
flow geometry and the nature of the spectral state. From the fitted result, we get some signature of iron line with 
accretion rate, which also has been discussed. Finally, in \S 4, we present a brief discussion 
of our results and make concluding remarks.

\section{Observation and Data Analysis}
\label{sec:1}
In the present manuscript, we analyze both Swift/XRT and NuSTAR satellite observations of GX~339-4
BHC during its 2013 outburst to obtain TCAF parameters and to check signatures of double peaked Iron line.
\subsection{Swift}
\label{sec:2}
In our present analysis we use 0.8-9.0~keV Swift/XRT observation of GX~339-4 during 2013 outburst the timings of which match with the NuSTAR observations presented below.
The observation IDs are 32490015, 80180001, 80180002, 32898013, and 32988001. We use {\it xrtpipeline v0.13.2}
task to extract the event fits file from raw XRT data. All filtering tasks are done by {\it FTOOLS}. 
We analyze XRT Windowed Timing mode data for all the observation. To generate 
source and background spectra we run {\it xselect} task. We use 
swxwt0to2s6\_20130101v015.rmf file for the response matrix. We rebinned the spectra following the same way 
as of NuSTAR spectra.
\subsection{NuSTAR}
\label{sec:3}
We analyze five NuSTAR observations (marked as X02, X04, X06, X08 and X10, where 
X=800010130 is the initial part of the observation Ids) of BHC GX~339-4. 
NuSTAR consists of two independent grazing incidence telescopes which focus 
X-rays onto corresponding focal planes consisting of Cadmium Zinc Telluride 
(CZT) pixel detectors. NuSTAR is sensitive to X-ray energies from 4-79 keV 
and provides high spectral resolution at energies above 10~keV. The two focal 
planes are referred to as focal plane module (FPM) A and B. NuSTAR data were 
extracted using the standard NUSTARDAS v1.3.1 software. Each FPM is at the focus of a 
hard X-ray telescope with a focal length of 10.14~m. We processed the focal plane module A data.
We run `nupipeline' task to produce cleaned event lists  and `nuproducts' for light curve
and spectral file generation. We use $30^"$ radius region for both the source extraction and the background 
using ``ds9". We group the data using ``grppha" command, where we group the whole energy bin with 8 
bins in each group. For spectral fitting we use XSPEC \cite{arnaud96} version 12.8.1. To
fit the spectra with TCAF model in XSPEC, we have a TCAF model generated {\it fits} file,
using theoretical spectra generated by large number of sets of input parameters in modified CT95 code
and included it in XSPEC as an additive table model. These parameters are: $i)$ mass of the black 
hole ($M_{BH}$) in solar mass ($M_\odot$) unit, $ii)$ disk rate ($\dot{m_d}$ in Eddington rate, 
$\dot{M}_{Edd}$), $iii)$ halo rate ($\dot{m_h}$ in $\dot{M}_{Edd}$), $iv)$ size of the Compton cloud 
(i.e., shock location $X_s$ in Schwarzschild radius $r_g$), and $v)$ density jump in advective flow 
at Compton cloud boundary (i.e., compression ratio $R$ of the shock). These accretion rates collectively
carry information of viscosity inside. Thus we do not need to add viscosity as a separate parameter. This is also a 
fundamental property of the TCAF model. We use hydrogen column density 5$\times$~10$^{21}$~atoms~cm$^{-2}$ 
throughout the analysis as we consider the same mass used in \cite{dd15a}, throughout the analysis. 
All the errors are calculated at $1\sigma$ confidence level.

\section {Results of model fit}
\label{sec:4}
We study Swift/XRT and NuSTAR observations with TCAF solution based fits file which used 
four physical disk parameters, as we froze the mass at $5.8M_\odot$. 
TCAF solution fitted results are given in Table~1. In Col. 2, we show the MJD of    
observations, and in Cols. 3, 4, 5 and 6, TCAF fitted parameters such as the disk rate, the halo rate, the 
shock location and the compression ratio are written. 
Column 13 gives $\chi^2$ and degrees of freedom. Other Columns are related to the iron lines and discussed below.

\subsection{Accretion geometry from TCAF fit}
\label{sec:5}
High resolution of Swift and NuSTAR data provides us with a better opportunity to study spectral and line 
properties using TCAF. In earlier papers, details about TCAF fits have been mentioned and they are not repeated here 
(\cite{dcm14,dd15a,dd15b,mdc14,molla16,jana16,chat16}). %(Debnath et al. 2014, Mondal et al. 2014; Debnath et al. 2015a, 2015b; Jana et al. 2016). 
For all observations we use 0.8-9.0 keV Swift/XRT and 4-79~keV NuSTAR data. 
In Fig. 1(a-d), we present TCAF fitted parameters in the present context. In Fig. 1a, we give 
variations of the accretion rates of the Keplerian (lower curve; online red circles) and
sub-Keplerian (upper curve; online blue circles) components. In these cases the source was in the hard state. 
In Fig. 1b, we present the variation of the size of the Compton cloud which is the post-shock region of the
sub-Keplerian component \cite{c89}. In Fig. 1c, we show the shock strength, i.e., the ratio 
of the post-shock to pre-shock flow density experienced by the
low angular momentum component. This basically carries the information about the
density and optical depth inside the Compton cloud. In Fig. 1d, we show the variation of energy spectral index,
which clears the hard state of the system throughout. At the beginning of the outburst (MJD=56515.99), 
the shock was far away from the source and the shock strength was also high ($\sim 4.04$). In the final observation 
(Id=X10) disk accretion rates were down after $\sim 44$ days of X08, indicating the end of the outburst. 
This value is higher than that on X04 day, although relative ratio (halo rate/disk rate) of the accretion rates
is higher in the final observed day.

\begin{figure}
\centerline{
\includegraphics[scale=1.,angle=0,width=8.truecm]{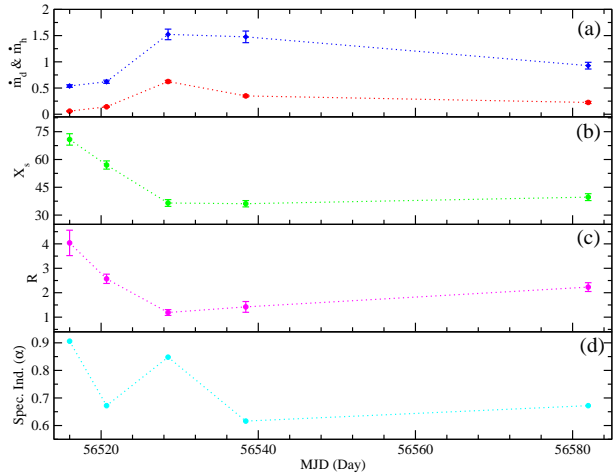}
}
\caption{Variation of (a) TCAF model fitted disk and halo rate, where halo rate is always higher than
disk rate, (b) location of the shock ($X_s$) with day and (c) shock strength, during the 2013 outburst of 
GX~339-4, using Swift and NuSTAR data. In bottom panel (d) variation of energy spectral index ($\alpha$) with day.} 
%\label{kn: fig1}
\end{figure}

In the first (X02) observed day, the disk rates are less compared to those on other days as the 
source was in the initial rising phase of the outburst. 
On X02, when the outburst started, matter was moving with viscous timescale and started to form a 
Keplerian disk by gradually increasing its rates. The flow was always fully dominated by 
the sub-Keplerian rate as fresh matters were coming with low angular momentum and 
since the Compton cloud could not be cooled down, states were always hard. At some point in time,
 viscosity started going down, and Keplerian disk rate also started declining. 
As matter supply is reduced, the disk rate became lower also. The whole system was dominated by 
sub-Keplerian flows throughout.

When it comes to the evolution of the size of the Compton cloud, it is to be remembered that the 
outer boundary of the Compton cloud is the shock itself. We find that the shock started to 
move towards the black hole from a location $X_s=70.83~r_g$ to $\sim 36~r_g$ 
(where, Schwarzschild radius $r_g=2GM_{BH}/c^2$) within $\sim 24$~days on $4^{th}$ observation (X08 on MJD=56538.41).
On the last observation (X10 on MJD=56581.99), the shock again moved away and observed at $\sim 39.68~r_g$ (see, Fig. 1b). 
The shock compression ratio $R$ (=$\rho_+ / \rho_-$, where $\rho_+$ and $\rho_-$ are post- and pre-shock densities 
respectively) was always observed to be higher than unity and became weaker on X06 and X08 observation days.
On those days shock was also closer to the black hole. Interestingly, in the whole outburst the shock is 
within 100 $r_g$, in agreement with the results of earlier authors.
After that for $\sim 44$ days there were no observation. It is possible that the source 
covered other spectral states, such as hard-intermediate, soft-intermediate and 
soft states but we may have missed them. On the X08 observation day, the shock was 
found to be receding and becoming strong again. High values of both $R$ 
and $\dot{m_h}$ indicate that the source was in the hard state during the entire observations. Due to the absence of 
soft state, we call this as a ``failed outburst". These results are in agreement with \cite{furst15}.

\subsection{Spectral analysis using TCAF}
\label{sec:6}
For the first time we fit the Swift and NuSTAR observation with TCAF model. As the Swift and NuSTAR have good spectral
resolution, data fits well with TCAF solution. In Fig. 2a, unfolded spectra of all the observations 
after fitting with TCAF and LAOR model \cite{laor91} of double peaked iron line.
From the TCAF model fitted parameters we see that the halo accretion rate is always greater than the 
disk accretion rate. Thus we infer that the source was in the hard state during this particular outburst.
In Fig. 2b, we show the residuals of these fits. Due to the characteristic shape of the double peaked
iron line and its very interesting variation with time, we show this 
separately in Fig. 3. Table 1, gives the fitted parameters from TCAF and LAOR models.
The inclination angle was chosen to be $60^o - 66^o$ (Col. 12) throughout. The emitting region sizes 
($R_{in}$ and $R_{out}$) are also given in the Table. Note that, the inclination angle chosen by us
is within the range predicted by other workers discussed in the Introduction. The inner edge of the
emitting region gives us the lower limit of the marginally bound orbit. The lowest of $R_{in}=1.21$ corresponds to
the marginally bound orbit of a Kerr black hole of $a=0.99$ \cite{shap83}. 
\begin{figure*}
\centerline{
\includegraphics[scale=1.,angle=270,width=7.0truecm]{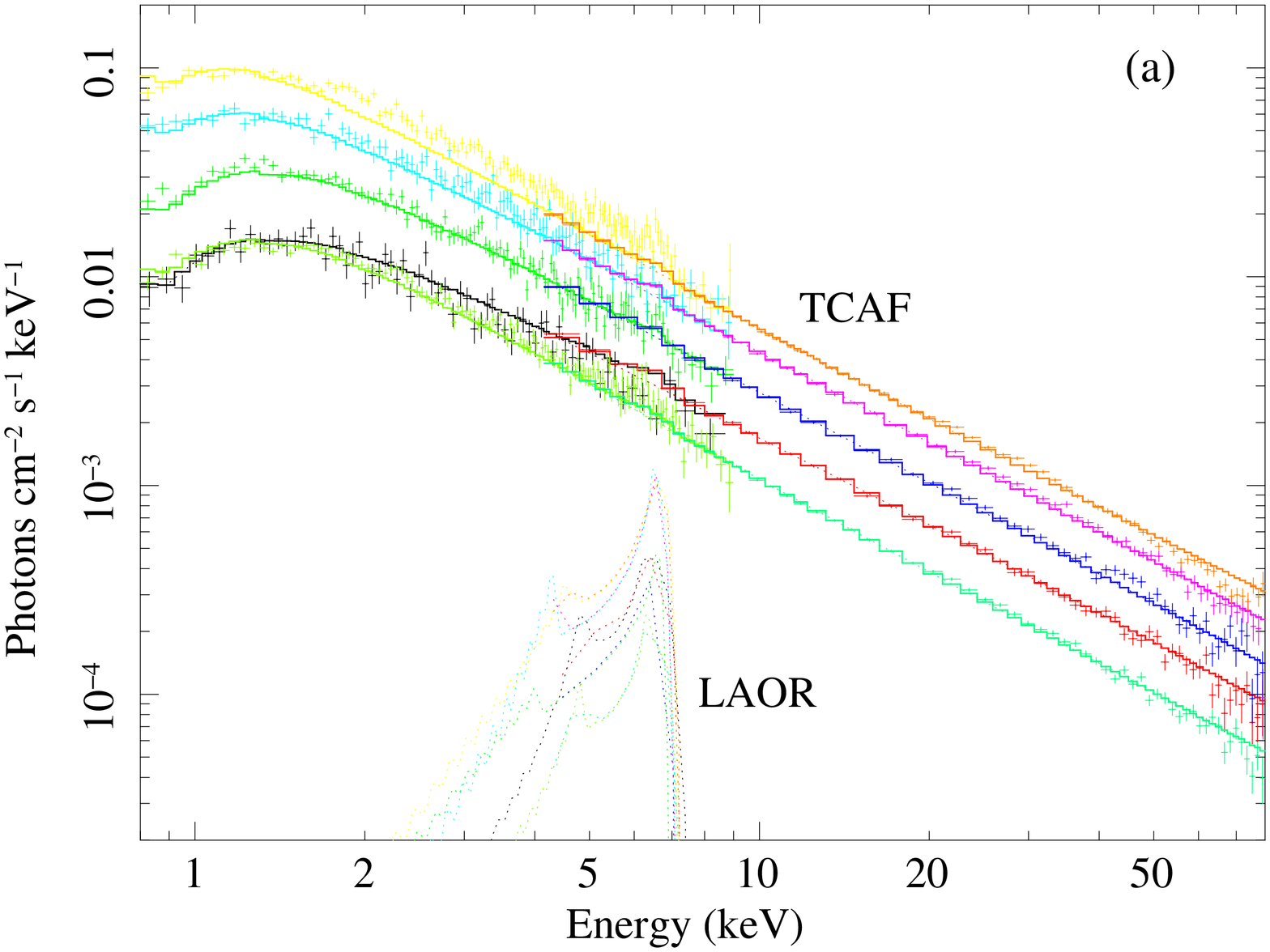}
\includegraphics[scale=1.,angle=270,width=7.0truecm]{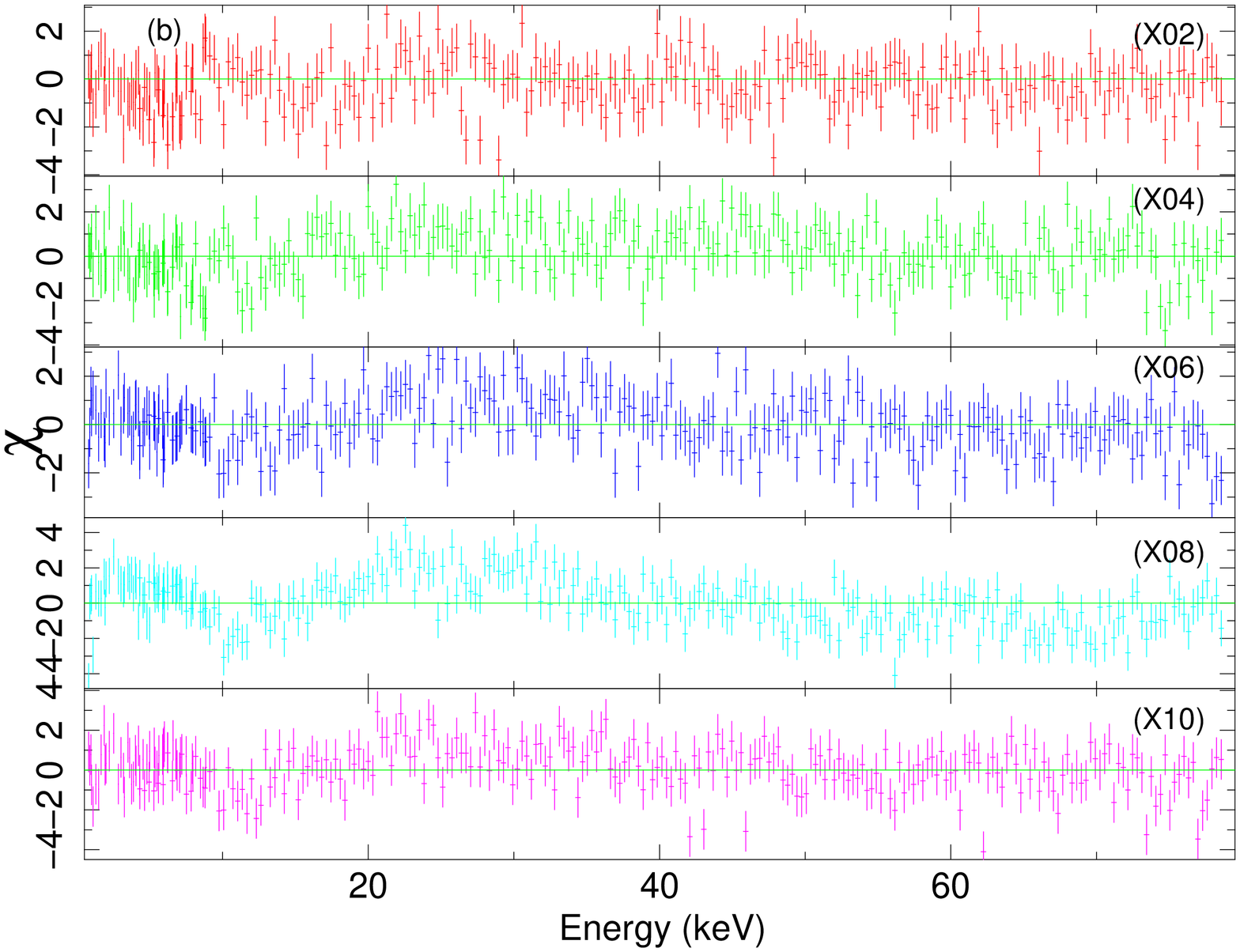}
}
\vskip0.7cm
\caption{In (a) unfolded spectra of all the observations fitting
with the TCAF model and the LAOR model of double peaked iron line,
(b) TCAF and LAOR model fitted residuals for the observations X02-X10 using Swift and NuSTAR data. 
}
\end{figure*}

\begin{figure*}
\centerline{
\includegraphics[scale=1.5,angle=0,width=8.0truecm]{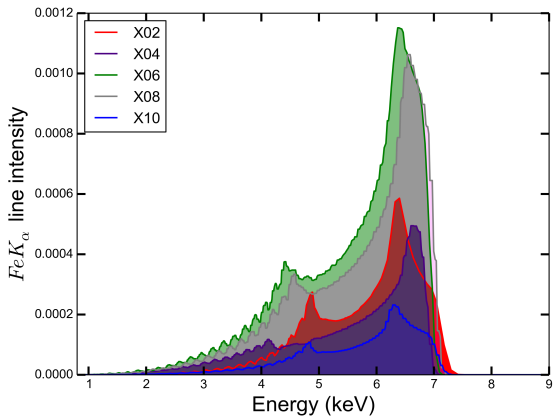}
}
\vskip0.7cm
\caption{Variation of $Fe~K_\alpha$ line intensity of all the observations fitting
with the TCAF model and the LAOR model. This variation confirms the signature of double horn line.
}
\end{figure*}

\section{Concluding Remarks}
\label{sec:7}
We have analyzed Swift/XRT and NuSTAR spectra of a very well-known Galactic stellar mass black hole source 
GX~339-4 during its 2013 outburst using TCAF solution along with LAOR model. 
We find that as the day progresses, both the accretion rates ($\dot{m_d}$ 
and $\dot{m_h}$) increase, the shock moves in (i.e., the Compton cloud slowly collapses due to rapid cooling) 
and its strength decreases. We observe that the source was in the hard state as the halo 
rate is always greater than the disk rate and the energy spectral index ($\alpha$ in Col.7) 
is $<$1 for all the observed days. 

According to \cite{ct95} and \cite{ebisawa96}, an outburst 
is triggered due to a sudden rise in viscosity at the outer edge of the disk and is 
turned off when this viscosity is reduced. If viscosity does not rise above a 
critical value, the high angular momentum and highly viscous cooler 
Keplerian disk component does not form \cite{c90} and as a result, we may miss the softer states, 
such as the soft and soft-intermediate spectral states. On the other hand, the most plausible `gap' where these
states could have been seen is that between X06 and X08, which appears to be impossible for this 
object \cite{nowak99,tomsick08,d10,nandi12}. %(see, Nowak et al. 1999; Tomsick et al. 2008; Debnath et al. 2010; Nandi et al. 2012). 
So, it is likely that the viscosity did not rise above the critical value to have a strong Keplerian component.
Furthermore, we find that the residual requires a double peaked iron line produced by a large
rotational velocity in all the days, characteristics of the inner region of the flow.
The double peak is clearly due to rotation, as the highest
intensity of the lines are observed during the highest accretion rate days with relatively smaller
size of the Compton cloud.
We also observed a significant change in line intensity during the days when the accretion rate 
is increased and the shock was closer to the black hole (Col. 10 of Table~1). In Col. 9, Laor model 
fitted power-law emissivity index $q$ is shown. This index indicates different regions of the disk follow different
emissivity laws. 
The present TCAF ${\it fits}$ file does not include the Iron line and the general relativistic effects. 
Thus we invoked `LAOR model' as a phenomenological model to understand the line emission profile and its
time variation. According to \cite{laor91}, double-peaked profile is generally produced from the
outer part of the disk and the red shifted component dies out at regions very close to the black hole. 
Based on the present fitted parameters, though not very self-consistently obtained from TCAF solution alone,
we find the upper limit of Kerr parameter $a$ of GX 339-4 to be $\sim 0.99$
and the disk inclination angle is in between $60^o - 66^o$.
In future, we will include these effects and hope to fit the data with TCAF only.

\begin{table}
%\footnotesize
\scriptsize
\begin{center}
\caption{0.8-9.0 keV and 4-79 keV TCAF+LAOR Model Fitted Parameters}\label{Table 1}
\begin{tabular}{lccccccccccccc}
\hline
Id&Day&$\dot{m_d}$&$\dot{m_h}$&$X_s$&R&$\alpha$&$R_{in}$&$q$&LD&$R_{out}$&${\it i}$&$\chi^2$/DOF\\
  &(MJD) &    &       &($r_g$)  & & & &  &  &  \\
%  &(MJD) &($\dot{M}$$_{Edd}$)  &($\dot{M}$$_{Edd}$)&($r_g$)  & & & &  &  &  \\
(1)& (2) & (3) &(4)& (5) & (6) &  (7) & (8) & (9) & (10) &  (11)  &(12) &(13) \\
\hline

X02&D15.99&$0.059^{\pm0.009}$&0.538$^{\pm0.025}$&$70.83^{\pm3.08}$&$4.04^{\pm0.52}$&$0.906$&5.53&2.01&8.64E-4&61.75&64.95&766.61/566  \\
X04&D20.70&$0.142^{\pm0.011}$&0.620$^{\pm0.030}$&$57.01^{\pm2.21}$&$2.57^{\pm0.19}$&$0.672$&1.24&2.10&6.61E-4&16.16&62.59&917.35/807  \\
X06&D28.52&$0.624^{\pm0.023}$&1.523$^{\pm0.101}$&$36.52^{\pm1.84}$&$1.19^{\pm0.12}$&$0.848$&1.24&1.79&1.82E-3&30.52&64.30&784.98/739  \\
X08&D38.41&$0.349^{\pm0.018}$&1.476$^{\pm0.111}$&$36.13^{\pm1.71}$&$1.42^{\pm0.22}$&$0.616$&1.21&1.77&1.77E-3&46.94&61.50&1423.99/848 \\
X10&D81.99&$0.225^{\pm0.020}$&0.927$^{\pm0.064}$&$39.68^{\pm1.85}$&$2.23^{\pm0.18}$&$0.672$&1.24&1.73&3.17E-4&44.85&65.00&1067.64/875 \\
\hline
\end{tabular}
\leftline {Here X=800010130 and D=565, are the initial part of the observation Ids. and MJDs.}
\leftline {$\dot{m_h}$, and $\dot{m_d}$ represent TCAF model fitted sub-Keplerian (halo) and Keplerian (disk) rates 
in} 
\leftline{Eddington rate respectively. $X_s$ (in Schwarzchild radius $r_g$), and $R$ are the model fitted shock} 
\leftline{location and compression ratio values respectively. ${\it i}$, is the disk inclination angle.}
\leftline {LD and $q$ represent Laor model fitted line strength and power-law emissivity index.} 
\end{center}
\end{table}

\begin{acknowledgements}
We are thankful to Dr. Varun Bhalerao for the tutorial of NuSTAR analysis during RETCO-II
conference. We are also thankful to the anonymous referee for useful suggestions to improve the quality of the manuscript. 
SM acknowledges support from a post doctoral grant of the Ministry of Earth Science, 
Govt. of India and FONDECYT post-doctoral fellowship grand (\# 3160350). 
DD acknowledges support from the project fund of DST sponsored Fast-track 
Young Scientist (SR/FTP/PS-188/2012). This research has made use of the
NuSTAR Data Analysis Software (NuSTARDAS) jointly developed by the ASI Science
Data Center (ASDC, Italy) and the California Institute of Technology (Caltech, USA),
and the XRT Data Analysis Software (XRTDAS) developed under 
the responsibility of the ASI Science Data Center (ASDC), Italy.
\end{acknowledgements}

\end{document}